\documentclass[twocolumn,preprintnumbers,amsmath,amssymb]{revtex4}
\usepackage{graphicx}
\usepackage{graphicx}
\usepackage{dcolumn}
\usepackage{latexsym}
\usepackage{amssymb}
\usepackage{amsfonts}
\usepackage{amsmath}

\begin{document}

\title{A Diffusive Strategic Dynamics for Social Systems}

\author{E. Agliari}
\affiliation{Dipartimento di Fisica, Universit\`{a} degli Studi di Parma,
viale Usberti 7/A, 43100 Parma, Italy}
\affiliation{Theoretische Polymerphysik, Universit\"{a}t Freiburg,
Hermann-Herder-Str. 3, D-79104 Freiburg, Germany}
\author{R. Burioni}%
\affiliation{Dipartimento di Fisica, Universit\`{a} degli Studi di Parma,
viale Usberti 7/A, 43100 Parma, Italy}
\affiliation{INFN, Gruppo collegato di Parma, Viale Usberti 7/A, 43100 Parma,
Italy}
\author{P. Contucci}%
\affiliation{Dipartimento di Matematica, Universit\`a di Bologna, Piazza di Porta S. Donato 5, 40126 Bologna, Italy
}

\begin{abstract}
We propose a model for the dynamics of a social system, which includes 
diffusive effects and a biased rule for spin-flips, reproducing
the effect of strategic choices. 
This model is able to mimic some phenomena taking place
during marketing or political campaigns. Using a cost function
based on the Ising model defined on the typical quenched
interaction environments for social systems (Erd\"{o}s-Renyi
graph, small-world and scale-free networks), we find, by numerical
simulations, that a stable stationary state is reached, and
we compare the final state to the one obtained with standard
dynamics, by means of total magnetization and magnetic
susceptibility. Our results show that the diffusive strategic
dynamics features a critical interaction parameter strictly lower
than the standard one. We discuss the relevance of our findings in
social systems.
\end{abstract}

\maketitle

\section{Introduction}
In the past few years the application of statistical mechanics to
social phenomena gave rise to interesting models, which were able
to capture some general mechanisms in opinion forming. In these
models, the relations between people in a group are represented by
a network with a given topology, where sites are people, links
model interactions and the opinion of a single agent is typically
represented by a discrete variable on the corresponding site.

One of the most important aspects in opinion forming within a community is the dynamics through which decisions of single agents take place, possibly leading the system to a stationary final state. In this framework different approaches have been considered, in order to investigate the configuration reached by the system at different time scales. A common approach, as in voter model and Axelrod dynamics (see e.g. \cite{castellano,liggett,axelrod}), is to introduce directly a dynamical rule according to which discrete variables, representing the choice of a single agent, evolve. Then, the average value of these discrete variables and its dependence on the dynamics itself as well as on the initial conditions is investigated. However, in many situations, ranging from polls to marketing analysis, relevant global parameters describing the behavior of large but finite subsamples of the populations are not  rapidly changing on the time scale considered. If the results of the experimental measurements are stable,  then it can be meaningful to analyze the social system by looking to its equilibrium  or stationary behavior. This approach, directly linked to statistical mechanics and used in social sciences,  is based on the introduction of a dynamics derived from a cost function $H$ \cite{durlauf,galam,bouchaud,gallo}, which depends on the configuration of the system and on a set of parameters, measuring the interaction between people in the community. The cost function theory has also shown its potentials in micro-economy especially
thanks to the work of Mc Fadden \cite{mcfadden}. There, quantitatively precise predictions were made on social behavior by means of {\it discrete choice theory}, a simple probabilistic approach based on the individual independent choice assumption, which corresponds to non-interacting agents. 
Among the main purpose of the present paper there is the attempt to extend to interacting systems the approach pioneered in 
\cite{mcfadden} and successfully applied to social systems by Durlauf \cite{durlauf} within the framework of equilibrium
statistical mechanics of deterministic systems. Our work here extends those previous findings in several directions: first
it assumes that there is interaction (including the random case) among the agents and investigate the possible steady states by means of natural dynamical systems usually encountered in social behavior. The dynamics is implemented with a cost function in which the parameters measuring the interactions between people can be assumed to be frozen. In real world phenomena the dynamics has two built-in time scales: one for the opinion flip, the other for the interaction change. In most of the social and economical examples the interactions among individuals change at a rate enormously slower than the opinions of each of them. In modeling such situations the interactions variables will be considered as frozen during the whole run of the dynamics and the physical quantities will be averages of stationary
states. The procedure is known as ``quenched" stationary state and parallels the standard quenched equilibrium measure of disordered systems (like spin glasses) in which the free energy is
computed averaging in the disorder after the logarithm.

Once the cost function and the related evolution dynamics are identified, the system can be studied numerically, for example via Monte Carlo simulations. In particular, if one expects that the final state reached by the system will obey an equilibrium Boltzmann distribution, the dynamics is usually implemented to lead the system to that state, imposing a detailed balance condition.
However, this  is quite unrealistic in real social communities.

As a matter of fact, dynamical evolution within social networks often exhibits a \textit{strategic} character:  each agent often chooses between his neighbors  and decides which link to activate at a given time, according to his own advantage, based on the observation of what happens to his neighbourhood.

A common situation is the following: people try to convince their neighbours in order to share the same opinion or trait because of ideological reasons or, also, because this translates in some
economical advantage. To fix ideas let us suppose that the opinion or the trait considered can be
described by a binary variable. For example, during an electoral campaign or before a
referendum, people try to convince their acquaintances to support a given candidate or a given
position. In a different context we can think about a community of people where each agent has
made a subscription to a phone company. Let us suppose that only two different companies,
A and B, exist so that we can distinguish between A-users and B-users. Now, fares for
phone-calls are different according to whether the call occurs between customers of the same
company or between customers of different companies, being higher in the latter case. As
a result, for an A-user
(B-user) the optimal situation is when all his acquaintances are also
A-users (B-users) as he
can then enjoy low fares. Hence, each agent would like to induce his
neighbours to adopt
the same company.

In both situations cited above, strategies are possible if an agent
knows the neighbourhood of his acquaintances. As an example, if I want to increase the number of, say, A-users, among my acquaintances, I can either pick up a friend of mine randomly among those who are B-users and try to convince him to become an A-user, or select among my friends
the B-user whose acquaintances are mostly A-users. The latter strategy is of course
expected to be more effective.

Another important aspect in dynamics concerns the rule according to which agents are selected and given the possibility to change opinion. Of course, a deterministic updating, though computationally efficient, is unrealistic: the dynamics must contain a
stochastic character, reproducing the randomness typical of social interactions \cite{kossinets,ebel,palla,onnela}. More likely, we can assume that the opinion updating results from phone calls, mail exchanges and other contacts among agents and this exhibits a \textit{diffusive} feature. Hence, the newest updated agent will choose among his neighbours the next agent to be updated.

In this paper, we intend to study the equilibrium reached by the
system, where an explicit cost function has been introduced,
endowed with a dynamics which takes into account these two
aspects: strategy and diffusion. We first focus on the dynamical
process, modelling  the quenched social network by an
Erd\"{o}s-Renyi random graph. This graph provides a stochastic
network, able to capture some aspects of a real community, and
allowing for some exact calculations \cite{newman,bovier,ABC2008}.
We then extend our results to scale free and small-world networks,
which are known to reproduce the typical  topological features of
real social networks \cite{barabasi,SW,NWB,guido}.  We adopt as a
cost function the ferromagnetic Ising Hamiltonian which, being one
of the simplest model mimicking interactions amongst the agents of
social systems, allows us to focus on the dynamical process. In
particular, the interaction parameter $J$ here represents the
``imitation strength" and it measures how important it is for two
nearest-neighbours to agree. For instance, in the phone companies
example, a large value of $J$ corresponds to a situation where a
phone call between A-A or B-B users is much cheaper than a phone
call between A-B users. We focus here on ferromagnetic interactions
as these are considered to be the predominant feature of social interactions
in several contexts \cite{bond}.  Our dynamics can however be extended 
to the case of  more complex
interaction patterns.

We find that after a suitable relaxation time the values of the global observables of the system
display time averages independent of the initial
conditions and whose fluctuations decrease with the system size, indicating that a stationary situation is reached. We also recover the phase diagram expected for the Ising model on the Erd\"{o}s-Renyi random graph, small-world and scale-free networks, and this implies that  in all these cases there exists an interaction parameter $J_c$ such that if $J<J_c$  the number of A-users equals, on the average, the number of B-users, while if $J>J_c$ a prevailing group emerges. However, we evidence a remarkable difference: With respect to the case of a non-strategic dynamics, the critical region is always shifted to a lower value of the
interaction parameter $J$. In the example of the competition between the two phone companies, this means that, once the price policy has been set by the companies, i.e. once a given $J > J_c$ has been fixed by the price differences, if the equilibrium is reached by a strategic dynamics, the extent of the prevailing community is larger. The existence and the position of a transition point represent a key information in a social system, as they signal an unstable situation, which is to be favored or avoided, depending on the meaning of the global parameter. Moreover, it is important to understand what influences the position of the critical region in order to control and tune it, if possible. Our result suggests
that the onset of the critical regime  is indeed influenced by the dynamics.

The paper is organized as follows: in Sec.~\ref{sec:model} we present our model and we describe the topologies it is embedded in. Then, Sec.~\ref{sec:dynamics} is devoted to the description of the strategic diffusive dynamics introduced and in Sec.~\ref{sec:numerics} we show our results.
Finally, Sec.~\ref{conclusions} is left for conclusions and outlook.

\section{Model and notations}\label{sec:model}
A social network is meant as a (typically large) set of people or groups of people, also called ``agents'', with some pattern of interactions between them. This can be efficiently envisaged by means of a graph whose nodes represent agents and links between two of them represent the existence of a relationship (which could be acquaintanceship, friendship, etc.). Therefore, each agent $i$ is connected with a set of ``nearest-neighbours'', whose number $\alpha_i$ is referred to as the ``degree'' of the node $i$.

Now, several kinds of graph have been proposed in the past as models able to mimic the features displayed by a real population, and they are all built starting from three main topologies: random graphs, small-world and scale-free networks.
The random graph introduced by Erd\"{o}s and R\'{e}nyi (ER) \cite{erdos} is one of the most studied since it combines a stochastic character with an easy definition which allows to calculate exactly many interesting quantities \cite{ABC2008}. However, real social networks have been shown to feature some peculiar topological and metric properties, which are not all included in the ER random graph.
Two typical features are: power-law degree distribution (scale-free topology), shortness of geodetical path (small-world phenomenon).

The ER random graph can be defined as follows: given a number $N$ of nodes, we introduce connections between them in such a way that each pair of vertices $i,j$ has a connecting link with independent probability $p$. As a consequence, the probability $p_{\alpha}$ that a node in a random graph has degree exactly equal to $\alpha$ is given by the binomial distribution \cite{newman}. The small-world network can be built starting from a ring with $N$ nodes, and ``superposing" to the ring a random
graph with given mean connectivity $\alpha$, by adding a link to two points $i$ and $j$ on the ring with probability $p_{\alpha}=\alpha/N$ (but it is possible to obtain small-world topologies with different rules \cite{alainmartin}). To account for the large variability of the degree in real social networks, one of the most studied topologies is the scale-free graph, where the probability $p_{\alpha}$ that a node  has degree equal to $\alpha$ is given by a power law:
\begin{equation}
p_{\alpha} \sim {\alpha}^{-\gamma}
\end{equation}
with $1<\gamma<3$.

We now outline the general framework for modeling interactions among agents. First of all, we associate to each agent $i$ a binary variable $s_i = \pm 1$, representing the two possible forms of the considered opinion or trait. For example, $s_i = + 1$ might indicate that the $i$-th agent does support the current government or is an A-user, while $s_i=-1$ that he does not support the current government or that he is a B-user. The whole community, described by the set $\mathbf{s} = \{ s_1, s_2, ..., s_N \}$, will therefore be characterized by the mean value
\begin{equation}
m = \frac{1}{N} \sum_k s_k
\end{equation}
which can be measured by, say, a referendum vote or a survey.

We assume that agents do not possess any a priori bias towards $+1$ or $-1$ state, but they move towards a given trait as a result of the interaction with their nearest-neighbours. More precisely, we introduce a cost function $H_{ik}$ which quantifies the cost for individual $i$ to agree with individual $k$ as \cite{contucci}
\begin{equation} \label{eq:cost}
H_{ik}(s_i,s_k)=-J_{ik} s_i s_k,
\end{equation}
where $J_{ik}$ represents the strength of interaction between agents $i$ and $k$. When $i$ and $k$ agree ($s_i s_k =1$) we have a cost $H_{ik} = - J_{ik}$, while when they disagree ($s_i s_k =-1$) we have $H_{ik} = J_{ik}$. Thus, the interaction works in such a way that, when $J_{ik}>0$, then $i$ and $k$ tend to imitate each others assuming the same trait and vice versa when $J_{ik}<0$. The magnitude of $J_{ik}$ gives how important it is for $i$ to agree or disagree with $k$ \cite{bond}.

For the whole population we have the total cost function
\begin{equation}\label{eq:hamiltoniana}
H(\mathbf{s},\mathbf{J}) = \sum_{k \sim i} H_{ik} = - \sum_{k \sim i} J_{ik} s_i s_k,
\end{equation}
where the sum is extended over all the couples of nearest-neighbour agents denoted as $k \sim i$.

The cost function of Eq.~\ref{eq:hamiltoniana} is just the
well-known Ising Hamiltonian (see e.g. \cite{brush}) which can be
treated by statistical tools. As it is well known, the cost
function $H(\mathbf{s},\mathbf{J})$ does not lead to any natural
dynamics and it is a very interesting matter of investigation to
define a proper dynamics able to make the system evolve towards an
``equilibrium'' state. This
can be achieved in several ways: Apart from exact analytical
approaches, available only for special structure topologies (one
dimensional and two dimensional lattices) and mean-field
solutions, a number of approximate techniques have been developed,
including series expansions, field theoretical methods and
computational methods.

Here we adopt Monte Carlo (MC) numerical techniques in order to simulate the evolution of the system from a given initial configuration $\mathbf{s_0}$ to the stationary state, which in general does depend on the evolutionary dynamics $D$ we choose and on the parameters $J_{ik}.$

Notice that $\mathbf{J}$ can be chosen to be directed and group dependent and this may account, in the examples discussed in the introduction, for different influences and fares inter and intra different groups. For instance, if, say, the A company applies very high costs for phone calls between different users, we expect the pertaining interaction strengths to be very large. On the other hand, if for a B user costs for phone calls towards A users are only slightly more expensive then the relevant interaction strengths are small.

\section{Diffusive Strategic Dynamics}\label{sec:dynamics}
In order to simulate the evolution of the population described by the cost function in Eq.~\ref{eq:cost}, several different algorithms have been introduced. Among them a well-established one is the so-called single-flip algorithm which makes the system evolve by means of successive opinion-flips, where we call ``flip'' on the node $j$ the transformation $s_j \rightarrow -s_j$.

More precisely, the algorithm is made up of two parts: first we need a rule according to which select an agent to be updated, then we need a probability distribution which states how likely the opinion-flip is.

As for the latter, we adopt the well-known Glauber probability: Given a configuration $\mathbf{s}$, the probability for the opinion-flip on the $j$-th node is given by
\begin{equation} \label{eq:Glauber}
p(\textbf{s},j,\mathbf{J}) = \frac{1}{1 + e^{\displaystyle
 \Delta H(\textbf{s},j,\mathbf{J})  }},
\end{equation}
where $\Delta H(\textbf{s},j,\mathbf{J})$ is the variation in the cost function due to the flip $s_j \rightarrow -s_j.$
Notice that, for single-flip dynamics, the cost variation $\Delta H$
consequent to an opinion-flip only depends on the opinion of a few agents,
viz. the $j$-th one undergoing the flipping process and its $\alpha_j$
nearest-neighbours. This can be shown by spelling out the cost function
variation appearing in Eq.~(\ref{eq:hamiltoniana}):
\begin{equation}\label{eq:variation}
\Delta H (\mathbf{s},j,\mathbf{J}) = 2 s_j \sum_{i \sim j} J_{ij} s_i .
\end{equation}

Interestingly, as can be derived from Eq.~\ref{eq:Glauber}, each
opinion-flip is the result of a stochastic process featuring a
competition between an energetic and an entropic term: the lower
the cost of the opinion-flip and the more likely its occurrence. The external parameter $\mathbf{J}$ tunes the
probability for an energetically unfavourable event to happen: For
very low values of $\mathbf{J}$ any event is equally likely to happen
independently of the magnetic configuration, conversely,
for high values of $\mathbf{J}$, when the agent $j$ is surrounded by agents sharing the same opinion, the flipping of $s_j$ gets a rare event.

As already recalled, the opinion-flip probabilities just described can determine a
dynamics only after a prescription for updating the system has
been introduced. In other words, we first need a selection rule according to which extract agents, then the opinion of the selected agent will be possibly updated according to $p(\textbf{s},j,\mathbf{J})$. There exist several different choices for the first procedure, ranging from purely random to
deterministic.

Now, the most popular algorithms select nodes to be updated
according to a sequential order which, though computationally efficient,
appears rather artificial in a social network. Indeed, unless no predetermined strategies are at work, the random updating ($D=\mathcal{R}$) seems to be the most plausible. In this case the probability that the current configuration $\mathbf{s}$ changes into $\mathbf{s}_j'$ due to the flip $s_j \rightarrow - s_j$, reads
\begin{equation}\label{eq:prob_random}
\mathcal{P}^{\mathcal{R}} (\mathbf{s},j;\mathbf{J}) = \frac{1}{N} p(\mathbf{s},j,\mathbf{J}).
\end{equation}

The dynamics generated by $\mathcal{P}^{\mathcal{R}}$ has been
intensively studied in the past (see e.g. \cite{barkema}) and it
has been shown to lead the system to the usual equilibrium
(canonical) distribution, derived from the cost function $H$.

However, in a social context, we notice that an opinion-flip can occur as a result of a direct interaction (phone call, mail exchange, etc.) between two neighbours and if agent $i$ has just undergone an opinion-flip he will, in turn, prompt one out of his $\alpha_i$ neighbours to change opinion \cite{kossinets,ebel,palla,onnela}. This kind of picture can not be described by a random updating rule. Moreover, in many situations, the arbitrary agent $i$ aims to be surrounded by neighbours $j$ sharing his own opinion (being this a cultural trait or a phone subscription), i.e. $s_i s_j=1$, because this translates in some advantage for agent $i$.

Here we want to explore a relaxation dynamics, $D=\mathcal{S}$,
able to take into account these aspects, namely a realistic
selection rule and a proper strategy. With respect to traditional
dynamics, $\mathcal{S}$ displays two important features:
\textit{i}. the selection rule exhibits a \textit{diffusive
character}: The sequence of sites selected for the updating can be
thought of as the path of a random walk moving on the graph
representing the social network; \textit{ii}. the diffusion is
\textit{biased}: The $\alpha_i$ neighbours are not equally likely
to be chosen but, amongst the $\alpha_i$ neighbours, the most
likely to be selected is also the most likely to undergo an
opinion-flip, namely the one which minimizes $\Delta
H(\mathbf{s},j,\mathbf{J})$. This corresponds to a local strategic
choice of agent $i$, as the chance to obtain an opinion flip is
high, though the stochastic character is preserved.

Let us now formalize how $\mathcal{S}$ works. Our MC simulations are made up of successive steps, each of them follows as:

- Being $i$ the newest updated agent (at the very first step $i$ is extracted randomly from the whole set of agents), we consider the corresponding set of nearest-neighbours defined as $\mathcal{N}_i=\{ i_1, i_2, ..., i_{\alpha_i}\}$; we possibly consider also the subset $\tilde{\mathcal{N}}_i \subseteq \mathcal{N}_i$ whose elements are nearest-neighbours of $i$ not sharing the same opinion: $j \in \tilde{\mathcal{N}}_i \Leftrightarrow j \in \mathcal{N}_i \wedge s_i s_j = -1$. Now, for any $j \in \mathcal{N}_i$ we compute the cost function variation $\Delta H (\mathbf{s},j,\mathbf{J})$, see Eq.~\ref{eq:variation}, which would result if the flip $s_j \rightarrow -s_j$ occurred; notice that $\Delta H (\mathbf{s},j,\mathbf{J})$ involves not only the nearest-neighbours of $i$, but also its next-nearest-neighbours.

- We calculate the probability of opinion-flip for all the nodes
in $\mathcal{N}_i$, hence obtaining
$p(\textbf{s},i_1,\mathbf{J}),p(\textbf{s},i_2,\mathbf{J}),...,p(\textbf{s},i_{\alpha_i},\mathbf{J})$,
where $p(\textbf{s},j,\mathbf{J})$, see Eq.~\ref{eq:Glauber}, is
the probability that the current configuration $\textbf{s}$
changes due to a flip on the $j$-th site.

- We calculate the probability ${\cal P}^{\mathcal{S}}
(\textbf{s};i,j;\mathbf{J})$ that, among all possible $\alpha_i$
opinion-flips considered, just the $j$-th one is realized. This
probability is defined as:
\begin{equation} \label{eq:probability}
{\cal P}^{\mathcal{S}} (\textbf{s};i,j;\mathbf{J}) \equiv
\frac{p(\textbf{s},j,\mathbf{J})}{\displaystyle \sum_{k \in
{\mathcal{N}_i}} p(\textbf{s},k,\mathbf{J})},
\end{equation}
namely it follows by properly normalizing the
$p(\textbf{s},j,\mathbf{J})$. Notice that, according to
Eq.~\ref{eq:probability}, among the nodes included in
$\mathcal{N}_i$, those which are more likely to be flipped are
also the more likely to be selected for the spin-flip.

We can possibly restrict the choice just to the set
$\tilde{\mathcal{N}}_i$, hence defining $\tilde{{\cal
P}}^{\mathcal{S}} (\textbf{s};i,j;\mathbf{J}) \equiv
p(\textbf{s},j,\mathbf{J})/ \sum_{k \in {\tilde{\mathcal{N}}_i}}
p(\textbf{s},k,\mathbf{J})$. The probabilities ${\cal
P}^{\mathcal{S}}$ and $\tilde{{\cal P}}^{\mathcal{S}}$ are non
trivially different from each other, due to the fact that they
depend not only on the magnetic neighbourhood of the current site
$i$, but also on the next-neighbourhood. As a consequence, in
general, the next flip will not necessarily occur in
$\tilde{\mathcal{N}}_i$, not even for large coupling strength. Nonetheless, as we have verified numerically, the
most striking result induced by the strategic dynamics, i.e. a
shift of the critical region towards lower values of the
interaction parameter (vedi infra) is preserved if we adopt
$\tilde{{\cal P}}^{\mathcal{S}}$ instead of ${\cal
P}^{\mathcal{S}}$.

- According to the normalized probability ${\cal P}^{\mathcal{S}}$
(see Eq.~\ref{eq:probability}), we extract randomly the node $j
\in \mathcal{N}_i$ and realize the opinion flip $s_{j} \rightarrow
- s_{j}$.

- We set $j \equiv i$ and we iterate the procedure.

Some remarks are in order now.
The detailed balance condition usually implemented in standard dynamics and leading to a standard Boltzmann distribution, is explicitly violated
(see \cite{BBCV2002,ABCV2005} for more details).
This does not contradict our dynamic intent: the evolution is not meant to recover
a canonical Bolzmann equilibrium, but rather to model a realistic
dynamics making the system evolve.
By comparing Eq.~\ref{eq:prob_random} and Eq.~\ref{eq:probability}, we notice that the latter has an additional  site dependence. As a result, the analytical approach to the master equation is extremely difficult. Using the expressions for the probabilities in Eq.~\ref{eq:probability}, the evolution equation
for the probability $P(\textbf{s},i,t;\mathbf{J})$ that, at time $t$, the system is in the configuration $\textbf{s}$ and the last spin flip occurred at site  $i$ reads:
\begin{eqnarray} \label{eq:mastereq}
P(\mathbf{s},i,t+1;\mathbf{J})-P(\mathbf{s},i,t;\mathbf{J}) =  \\
\sum_{k \in {\mathcal{N}_i}}
{\cal P}^{\mathcal{S}} (\textbf{s};k,i;\mathbf{J}) P(\textbf{s'}_i,k,t;\mathbf{J}) -
{\cal P}^{\mathcal{S}} (\textbf{s};i,k;\mathbf{J}) P(\textbf{s},i,t;\mathbf{J}), \nonumber
\end{eqnarray}
where $\textbf{s'}_i$ represents the spin configuration which
differs from $\textbf{s}$ only for the value of $s_i$.
Interestingly, the additional dependence on the sites can be
interpreted in term of a biased random walker, located on site $i$
and implementing the dynamics by interacting with the spin system
\cite{BBCV2002}. Therefore the master equation can be seen to
describe, at a given time $t$, the evolution of the probability
for the spin variables to be in the configuration $\textbf{s}$ and
for the random walker to be located at site $i$. This makes the
analytic approach difficult, even in the simplest lattices.
Some general results, mainly numerical, have been obtained previously and in
different contexts; we now briefly review some basic facts, while for
more detailed information we refer the reader to
\cite{BBCV2002,ABCV2005}. Firstly, the diffusive dynamics is able
to eventually drive the system to a thermodynamically well-behaved
steady-state, characterized by the choice of the interaction coupling
$\mathbf{J}$ and by the underlying topology, and also to recover
the expected phase transition. However, the steady-state reached
differs from the expected canonical equilibrium in a non-trivial
way \cite{BBCV2002}. A particular evidence of this is
provided by the fact that the critical temperature (or,
similarly the critical coupling) obtained with the diffusive
dynamics is significantly larger than the one expected and such an
effect can not be accounted for by a simple rescaling of the
temperature \cite{BBCV2002}. Moreover, such results appear to be
robust, as they hold also on general finite dimensional structures
and for spin-$1$ models \cite{ABCV2005}.

Finally, it is worth underlying that such a diffusive dynamics is
intrinsically meant for finite systems. First of all this is
consistent with our intent to model the dynamics
within a community which is, indeed, finite. Moreover, on
finite, connected graphs without traps, a random walk is always
recurrent and visits each site. This ensures that if our Monte
Carlo simulation is run long enough, each agent is selected for
the spin-flip a number of times sufficient to obtain a series of
decorrelated states over which perform statistical averages.
Interestingly, the rate at which a given site is visited depends
non-trivially on its local magnetic environment as well as by the
temperature and by the network topology. In particular, as shown numerically
in \cite{agliari2,agliari3}, sites more likely to be selected for
the updating correspond to borders between clusters.

\section{Numerics}\label{sec:numerics}
As mentioned before, the analysis of the diffusive dynamics has been carried out mainly from the numerical point of view by means of extensive Monte Carlo simulations \cite{barkema}. Here we focus on the particular case of interaction parameters $J_{ik}$ independent of the particular couple of agents considered, i.e. $J_{ik} \equiv J$, for any $i,k$; this allows to highlight the role of the dynamics leading the system to a stationary state and to understand how it possibly affects such stationary state and the average trait $m^D$.

In the simulation, once the network has been defined, we place a
binary variable $s_i$ on each node $i$ and allow it to
interact with its nearest-neighbors. Once the external parameter
$J$ is fixed, the system is driven by the single-flip dynamics and
it eventually relaxes to a stationary state characterized by
well-defined properties. More precisely, for an ER random graph characterized by parameters $\alpha$ and $J$, after a suitable time lapse $t_0$  and for
sufficiently large systems, measurements of a (specific) physical
observable $x(\mathbf{s},\alpha,J)$ fluctuate around an average
value only depending on the external parameters $J$ and $\alpha$.

We also verified that, for a system of a given finite size $N$,
the extent of such fluctuations scales as $N^{-\frac{1}{2}}$ (see
also \cite{BBCV2002,ABCV2005}), as indicated by standard
statistical mechanics for a system in equilibrium. The estimate of
the a given observable $\langle x \rangle$ is therefore obtained
as an average over a suitable number of (uncorrelated)
measurements performed when the system is reasonably close to the
equilibrium regime. The estimate is further improved by averaging
over different realizations of the underlying random graph with
fixed $\alpha$. In summary,
\begin{equation}
\langle x \rangle_{\alpha,J} \equiv \mathbb{E} \left[
\frac{1}{M} \sum_{n=1}^{M} x(\mathbf{s}(t_n))\right] , \; \;
t_n=t_0+n\mathcal{T}
\end{equation}
where $\mathbf{s}(t)$ denotes the configuration of the system
at time step $t$ and $\mathcal{T}$ is the decorrelation parameter
(i.e. the time, in units of spin flips, needed to decorrelate a
given magnetic arrangement from the initial state); the symbol $\mathbb{E}$ denotes the average over different realizations of the graph.\\
In general, during a MC run in a given sample we find statistical errors
which are significantly smaller than those arising from the
ensemble averaging (see also \cite{szalma}).

\begin{figure}\label{fig:term}
\includegraphics[width=90mm, height=70mm]{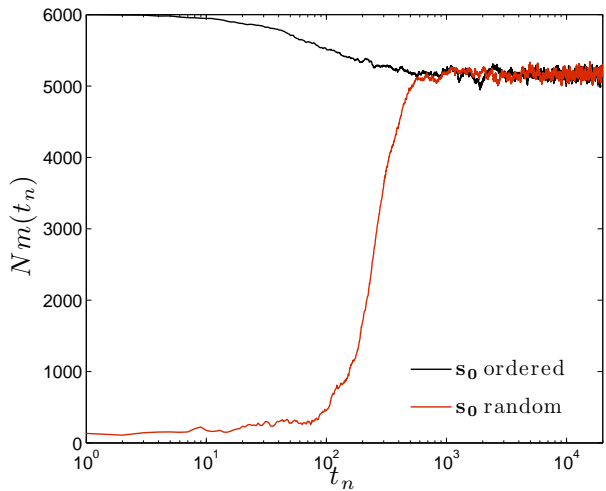}
\caption{Thermalization of a ER random graph made up of $N=6000$ agents and $\alpha = 30$. Two dramatically different initial configurations are considered and compared: an ordered configuration with $s_i=1$ for any $i$ (black) and a random configuration with $s_i=1$ ($s_i=-1$) with probability $1/2$ (red).}
\end{figure}

We stress once again that the final state obtained with the diffusive dynamics is stable, well-defined and, in particular, it does not depend on the initial conditions., i.e. it has all the properties of an equilibrium state. This is of course well-established for standard dynamics and it was also verified for our diffusive dynamics. An example is shown in Fig.~1 where, for a ER random graph, with given $(\alpha,J)$, the specific value around which $m(t_n)$ eventually fluctuates does not depend on the choice of the initial configuration selected for the simulation. To this aim we plotted $m(s(t_n))$ obtained starting with a completely ordered configuration ($m_0=1$) and with a completely random one. Moreover, we verified that the measurement of the (specific) observables within a macroscopic subsystem yields the same results characterizing the whole system. Similar results hold for scale free and small-world networks.

In the following we focus on systems of sufficiently large size so
to discard  effects related to small $N$. For the ER random graph, for a wide range of interaction constants $J$
and average coordinations  $\alpha$, we measure the average magnetization $\langle m
\rangle_{\alpha,J}$ (hereafter $\langle m \rangle$) and the susceptibility $\chi$, calculated as
\begin{equation} \label{eq:chi}
\chi(\alpha,J) \equiv  J N \left[ \langle m^2
\rangle - \langle m \rangle^2 \right].
\end{equation}
This quantity measures, at equilibrium, the reactivity of the system to a small external perturbation.
Moreover, we compare results obtained for our dynamics with those obtained through a well-established algorithm, i.e. the Glauber heath-bath with random updating ($D=\mathcal{R}$), which is known to lead the system to a canonical steady state.

\begin{figure}\label{fig:N6000}
\includegraphics[width=90mm, height=70mm]{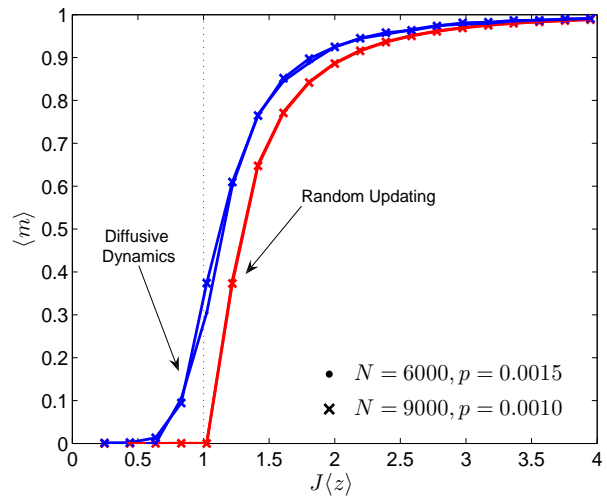}
\caption{Average opinion $\langle m \rangle$ for a population of $N=6000$ ($\bullet$) and $N=9000$ ($\times$) agents on a ER random graph with $p = 0.0015$ and $p = 0.0010$, respectively; the average number of nearest-neighbours is therefore the same for both systems, $\alpha=9$. Results obtained with a heath-bath dynamics ($\mathcal{R}$) and with the strategic dynamics ($\mathcal{S}$) are compared: for the latter a smaller critical parameter $J_c$ is found.}
\end{figure}

In Fig.~2 we show results pertaining to ER random graphs of different sizes ($N=6000$, $N=9000$), but keeping the average coordination number fixed ($\alpha= 9$ corresponding to $p=0.0015$ and $p=0.0010$, respectively). Their profiles display the typical behaviour expected for the Ising model on a random graph \cite{ABC2008} and, consistently with the theory, highlight a phase
transition at a well defined value $J_c^{\mathcal{S}}(\alpha)$. Otherwise stated, there exists a critical value of the parameter $J$ below which the system is spontaneously ordered.

Note however that $J_c^{\mathcal{S}}(\alpha)$ is appreciably
smaller than the critical value $J_c(\alpha) \approx 1/\alpha$
expected for the canonical Ising model defined on the ER random
graph \cite{ABC2008}.

\begin{figure}\label{fig:SF_SW}
\includegraphics[width=90mm, height=70mm]{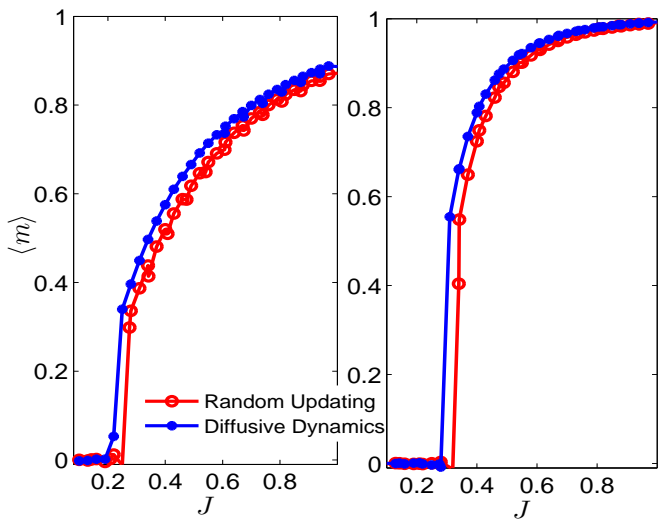}
\caption{Average opinion $\langle m \rangle$ for a population of
$N=1000$ agents on a scale-free (left panel) and small-world
(right panel) graph. Results obtained with a standard dynamics
($\mathcal{R}$) and the strategic dynamics ($\mathcal{S}$) are
compared: for the latter a smaller critical parameter $J_c$ is
found.}
\end{figure}

\begin{figure}\label{fig:N9000}
\includegraphics[width=90mm, height=70mm]{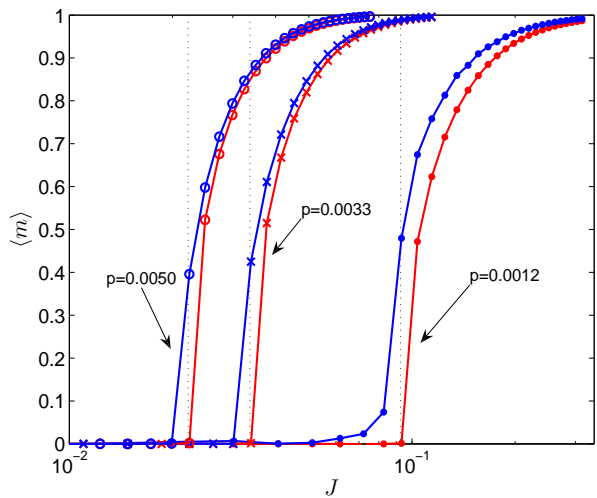}
\caption{Average opinion $\langle m \rangle$ for a population of
$N=9000$ agents with $\alpha = 11$ ($\bullet$), $\alpha=30$
($\times$) and $\alpha=45$ (open circles) on a ER random graph.
Results obtained with a standard dynamics ($\mathcal{R}$) and the
strategic dynamics ($\mathcal{S}$) are compared: for the latter a
smaller critical parameter $J_c$ is found.}
\end{figure}

An analogous behavior is also observed on scale-free and
small-world networks; data for the average opinion $\langle m
\rangle$ on these structures is shown in Fig.~3. More precisely,
for the scale-free network considered here, the degree
distributions for nodes follows the power law $p_{\alpha} \sim
\alpha^{-\gamma}$ with $\gamma=2.3$; as for the small-world network,
this has been built according to the prescriptions explained above
and in such a way that the average degree per node equals $4$.

Remarkably, similar diffusive dynamics have been shown to lead an
analogous decrease of the critical interaction parameter on
regular structures \cite{BBCV2002} and also for the spin-1 Ising
model \cite{ABCV2005,agliari2,agliari3}. In these cases it was
proved that a simple rescaling of the interaction constant $J$ can
not account for the differences between results produced by the
diffusive dynamics and a heath-bath dynamics. This feature
constitutes a first signature of the fact that the equilibria
generated by the diffusive dynamics are not governed by the
Boltzmann distribution.

As mentioned above, on an ER random graph the critical value $J_c$
depends on the system size and on the probability $p$, through
their product $\alpha$, i.e. $J_c \approx 1 / \alpha$. In order to
check if a similar behaviour also holds for the dynamics
$\mathcal{S}$, we now fix the size of the system and make $\alpha$
vary; results for $N=9000$ with $\alpha = 11,30,45$ are reported
in Fig.~4. Indeed, also for $J_c^{\mathcal{S}}$, we evidence a
monotonic increase with $\alpha$, however, in order to establish
the actual relation between $J$ and
$\alpha$, further extensive simulations are necessary.

Similar to what happens with the usual dynamics, the relaxation time needed to drive the system sufficiently close to the equilibrium situation is found to depend on the parameter $J$. More precisely, we experience the so called critical slowing down: the closer $J$ to its critical value, the longer the relaxation time.

We now turn to the susceptibility defined in Eq.~\ref{eq:chi};
results for the ER random graph are shown in Fig.~5. In the
thermodynamic limit, at the critical point $J_c$, the
susceptibility diverges, while for finite sizes the susceptibility
is expected to display a peak at $J_c$. This kind of behaviour is
found also when the diffusive dynamics is applied and $\chi$ just
peaks at $J_c^{\mathcal{S}}$. An important point is that the shape
of the curve is not modified, indicating that the reaction of the
system to an external perturbation is  conserved, with respect to
the usual equilibrium, in the vicinity of the critical point. The
social system is therefore expected to behave in the same way.
Hence, we have further evidence of the fact that the diffusive
strategic dynamics recovers the phase transition typical of the
Ising model, though providing a lower value for the critical
interaction parameter.

\begin{figure}\label{fig:chi}
\includegraphics[width=90mm, height=70mm]{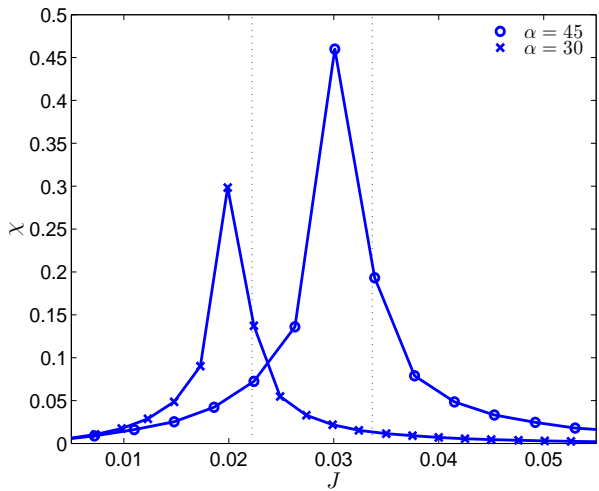}
\caption{Average susceptibility $\chi$ for a population of $N=9000$ agents with $\alpha=30$ ($\times$) and $\alpha=45$ ($o$) on a ER random graph obtained with the strategic dynamics. Notice that the function peaks at a value smaller than the critical $J_c$ expected for standard dynamics (dotted lines).}
\end{figure}

\section{Conclusions and Perspectives}\label{conclusions}

In this paper, we have introduced a dynamics for social systems displaying diffusive and strategic
character. This dynamics has been shown to relax
the system to thermodynamically
well-behaved steady states; in particular, after a suitable time,
the values of the global observables of the system
display averages independent of the initial
conditions (which can, at least, affect the orientation of the
asymptotic arrangement). The magnetization, representing the average trait reached by the system as a function of the interaction parameter $J$, features a transition at
a value of $J$ which is strictly lower than the one obtained with a non-strategic random choice for the opinion flips, on the same network. In particular,   on the ER random graph, also the shape of the susceptibility near the transition point is conserved, indicating that
the reaction of the social system to a small external perturbation in the stationary state is not modified.

This picture indicates that with a strategic (local) choice in opinion flips a full-consensus configuration is obtained for lower values of the interaction parameter $J$,
namely it is ``easier'' to obtain a community with an oriented opinion. Differently stated, in a society where a given value of the interaction $J$ is present, the number of people with an oriented opinion is higher if the equilibrium is reached by a strategic opinion flip. The shift effect on the critical parameter is a general feature of the strategic dynamics, and our results have been shown to hold on very different topologies, such as ER random graph, scale-free random networks with hubs and small-world networks \cite{barabasi}. Transition points in social systems lead to extreme sensitivity of the global parameter
to interactions, so its position and its shift represent a key information in the understanding of the stationary state. It can be interesting to enhance them or to avoid them, depending on the meaning of the
global parameter. 

Finally, our dynamics allows a number of generalizations concerning, for instance, the number of possible cultural traits or competitive companies admitted (namely the magnitude of the spin variable) \cite{ABCV2005}, the possible presence of an external magnetic field (representing the effect of external biases such as advertisement) or a more complicated set of constant parameters $J_{ij}$. In particular, $\mathbf{J}$ could be a directed, block matrix and this would account for different fares inter and intra distinct groups; in this case it would be very interesting to understand the conditions, in terms of $\mathbf{J}$ elements, for the realization of an oriented (i.e. magnetized) system.

\section*{Acknowledgment}
The authors are grateful to Adriano Barra for interesting suggestions and stimulating discussion. EA thanks the Italian Foundation ``Angelo della Riccia'' for
financial support.

\end{document}